\documentclass[aps,preprint,amsmath,amssymb]{revtex4}

\usepackage{graphicx}
\begin{document}

\title{Probe of anomalous neutrino couplings to W and Z in
medium energy setup of a beta-beam facility}

\author{A. B. Balantekin}
\email[]{baha@physics.wisc.edu}
\affiliation{Department of Physics,
University of Wisconsin, Madison, WI 53706, USA}

\author{\.{I}. \c{S}ahin}
\email[]{isahin@wisc.edu} \email[]{isahin@science.ankara.edu.tr}
\affiliation{Department of Physics, University of Wisconsin,
Madison, WI 53706, USA}
 \affiliation{Department of
Physics, Zonguldak Karaelmas University, 67100 Zonguldak, Turkey}

\author{B. \c{S}ahin}
\email[]{bsahin@wisc.edu} \email[]{dilec@science.ankara.edu.tr}
\affiliation{Department of Physics, University of Wisconsin,
Madison, WI 53706, USA}
\affiliation{Department of
Physics, Zonguldak Karaelmas University, 67100 Zonguldak, Turkey}

\begin{abstract}
Capability of medium energy setup of a beta beam experiment to probe
new physics contributions to neutrino-W and neutrino-Z couplings are
investigated. We employ the effective lagrangian approach of
Buchmuller and Wyler and obtain $95\%$ confidence level limits on
neutrino couplings to these gauge bosons without assuming the flavor
universality of the coupling of neutrinos. We show that a beta beam
facility with a systematic error of $2\%$ can place 10 times more
restrictive limit than present one on the deviations from the
electron neutrino-Z couplings in the Standard Model.
\end{abstract}

\maketitle

\section{Introduction}

Beta beams are electron neutrino and antineutrino beams
produced via the beta
decay of boosted radioactive ions  \cite{Zucchelli:2002sa} . Such decays
produce pure, intense and collimated neutrino or antineutrino
beams. In the original scenario ion beams are
accelerated in the proton synchrotron (PS) or super proton
synchrotron (SPS) at CERN up to a Lorentz gamma factor of $\gamma
\sim 100$, and then they are allowed to decay in the straight
section of a storage ring. Feasibility of this design has been
demonstrated in Ref. \cite{Autin:2002ms}. After the original
proposal, different options for beta beams were investigated.
A low gamma ($\gamma =5-14$) option was first proposed by Volpe
\cite{Volpe:2003fi}. Physics potential of low-energy beta beams was discussed
in detail. It was shown that such beams could have an important
impact on nuclear physics, particle physics and
astrophysics
\cite{McLaughlin:2003yg,Serreau:2004kx,McLaughlin:2004va,Volpe:2005iy,Balantekin:2005md,Balantekin:2006ga,Volpe:2006in,Jachowicz:2006xx,Bueno:2006yq,Lazauskas:2007va,Amanik:2007zy,Jachowicz:2008kx}.

Higher gamma options for the beta beams have also been studied in the
literature \cite{Volpe:2006in,BurguetCastell:2003vv,
BurguetCastell:2005pa, Migliozzi:2005zh, Donini:2006tt,
Terranova:2004hu, Huber:2005jk, Donini:2005qg,
Donini:2004hu,Mezzetto:2003ub}. A higher gamma factor provides
several advantages. Firstly, neutrino fluxes increase quadratically
with the gamma factor. Secondly, neutrino scattering cross sections
grow with the energy and hence considerable enhancement is expected
in the statistics. An additional advantage of a higher gamma option
is that it provides us the opportunity to study deep-inelastic neutrino
scattering from the nucleus. Very high gamma ($\sim 2000$) options
would require modifications in the original plan
such as using LHC and therefore extensive feasibility study
is needed. In this context medium energy setup is more appealing and
less speculative. We investigate the physics potential of a medium
energy setup ($\gamma =350 - 580$) proposed in
Ref. \cite{BurguetCastell:2003vv} to probe non-standard neutrino-$Z$
and neutrino-$W$ interactions. We do not make the {\it a priori} assumption of
the flavor universality of the coupling of neutrinos to these gauge bosons.

Neutrino-$W$ and neutrino-$Z$ couplings have been precisely tested
at CERN $e^{+} e^{-}$ collider LEP. Non-standard $W \ell \nu$
couplings are constrained via $W$ boson decay to leptons. It is
possible to discern neutrino flavor in $W^+\to \ell^+ \nu_{\ell}$
decay by identifying charged lepton flavor. Therefore individual
limits on neutrino-$W$ couplings for different neutrino flavors can
be obtained from the LEP data. On the other hand neutrino-$Z$
couplings are primarily constrained by the invisible $Z$ width, which receives
contributions from all neutrino flavors. Hence it is impossible to
discern possible universality violating neutrino-$Z$ couplings from
the LEP data alone. It is however possible to constrain new physics
contributions to $Z\nu\nu$ that respect universality. From the data
on $W^+\to e^+ \nu_{e}$ decay and invisible $Z$ width we set the
bounds of \cite{Yao:2006px}
\begin{eqnarray}
\label{LEPlimit}
-0.016\leq \Delta_e^{\prime} \leq 0.016\\
|\Delta_e+\Delta_\mu+\Delta_\tau|\leq 0.009
\end{eqnarray}
where the parameters $\Delta_e^{\prime},\Delta_e,\Delta_\mu$ and
$\Delta_\tau$ describe possible deviations from the SM coming from
new physics. They modify the charged and neutral neutrino current as
\cite{Masso:2002vj}
\begin{eqnarray}
\label{smcurrent} J^{CC}_\mu=\left[1+\Delta_e^{\prime}\right]\bar
{{\nu}_e}_{L}\gamma_\mu e_{L}\,,\,\,\,\,\,\,\,\,\,\,\,\,\,\,\,
J_\mu^{NC}=\frac{1}{2}\sum_{i=e,\mu,\tau}[1+\Delta_i]\bar{\nu}_{iL}
\gamma_\mu {\nu_i}_L
\end{eqnarray}
These new physics contributions respect universality of the coupling
of neutrinos to $Z$ if the equality
$\Delta_e=\Delta_\mu=\Delta_\tau$ holds. If we assume the
universality of the coupling of neutrinos to Z, LEP data give a
stringent limit of $-0.003< \Delta_e <0.003$.

On the other hand our purpose is to carry out a general treatment and we
do not {\it a priori} assume universality of the couplings of
neutrinos to gauge bosons. The processes isolating a single
neutrino flavor do not imply neutrino flavor universality and
therefore provide more information about new physics probes on
$Z\nu\nu$ couplings as compared to the invisible decay width experiments of
Z boson. There are experimental results from CHARM Collaboration
obtained from muon-neutrino and electron-neutrino scattering
reactions. We have the following limits from CHARM and CHARM II data
\cite{Vilain:1994gx,Dorenbosch:1986tb}
\begin{eqnarray}
\label{CHARMlimit} |\Delta_\mu|\leq 0.037 \, ,
\,\,\,\,\,\,\,\,\,\,\,\,\,\, -0.167\leq \ \Delta_e \leq 0.237 .
\end{eqnarray}
The plan of this paper is as follows: In the next section we outline the effective
Lagrangian approach. In section III we summarize the neutrino fluxes and the
cross sections for elastic, inelastic and deep-inelastic scattering and present our
main results. Finally Section IV includes concluding remarks.

\section{Effective Lagrangian for $Z\nu\nu$ and $W\ell\nu$ couplings}

There is an extensive literature on non-standard interactions of
neutrinos
\cite{Davidson:2003ha,Bell:2005kz,Mohapatra:2005wg,Mohapatra:2006gs,de
Gouvea:2007xp,Perez:2008ha,Balantekin:2008rc,Balantekin:2008ib}. New
physics contributions to neutrino-$Z$ and neutrino-$W$ couplings can
be investigated in a model independent way by means of the effective
Lagrangian approach. The theoretical basis of such an approach rely
on the assumption that at higher energies beyond where the Standard
Model (SM) is tested, there is a more fundamental theory which
reduces to the SM at lower energies: The SM is assumed to be an
effective low-energy theory in which heavy fields have been
integrated out. Such a procedure is quite general and independent of
the new interactions at the new physics energy scale.

We consider the $SU(2)_L\otimes U(1)_Y$ invariant effective
Lagrangian introduced in Ref. \cite{Buchmuller:1985jz}. Possible
deviations from the SM that may violate flavor universality of the
neutrino-V (V=Z,W) couplings are described by the following
dimension-6 effective operators:
\begin{eqnarray}
\label{eop1}
O_j=i(\phi^\dagger D_\mu \phi)(\bar \psi_j \gamma^\mu \psi_j)\\
O^\prime_j=i(\phi^\dagger D_\mu \vec{\tau} \phi)\cdot(\bar \psi_j
\gamma^\mu \vec{\tau} \psi_j) \label{eop2}
\end{eqnarray}
where $\psi_j$ is the left-handed lepton doublet for flavor
$j=e,\mu$ or $\tau$; $\phi$ is the scalar doublet; and $D_\mu$ is
the covariant derivative, defined by
\begin{eqnarray}
D_\mu=\partial_\mu+i\,\frac{g}{2}\,
\vec{\tau}\cdot\vec{W}_\mu+i\,\frac{g^\prime}{2}\,YB_\mu .
\end{eqnarray}
Here $g$ and $g^\prime$ are the $SU(2)_L$ and $U(1)_Y$ gauge
couplings, $Y$ is the hypercharge and the gauge fields $W^{(i)}_\mu$
and $B_\mu$ sit in the $SU(2)_L$ triplet and $U(1)_Y$ singlet
representations, respectively.

The most general $SU(2)_L\otimes U(1)_Y$ invariant Lagrangian up to
dimension-6 operators, containing new physics contributions that may
violate universality of the neutrino-V couplings, is then given by
\begin{eqnarray}
\label{lag}
{\cal L}={\cal
L}_{SM}+\sum_{j=e,\mu,\tau}\frac{1}{\Lambda^2}(\alpha_j\,O_j+\alpha^\prime_j
O^\prime_j)
\end{eqnarray}
where ${\cal L}_{SM}$ is the SM Lagrangian, $\Lambda$ is the energy
scale of new physics and $\alpha_j$, $\alpha^\prime_j$ are the
anomalous couplings. After symmetry breaking, Lagrangian in Eq. (\ref{lag})
reduces to \cite{Buchmuller:1985jz}
\begin{eqnarray}
{\cal L}^\prime=\frac{g}{\sqrt {2}}\left(J^{CC}_\mu
W^{+\mu}+J^{CC\,\dag}_\mu\,
W^{-\mu}\right)+\frac{g}{\cos\theta_W}\,J^{NC}_\mu\,Z^\mu ,
\end{eqnarray}
where $J^{CC}_\mu$ and $J^{NC}_\mu$ are charged and neutral
currents. They are given by
\begin{eqnarray}
\label{current1}
J^{CC}_\mu=\left[1+2\alpha^\prime_j\frac{v^2}{\Lambda^2}\right]\bar
{{\nu}_j}_{L}\gamma_\mu {\ell_j}_{L}
\end{eqnarray}
\begin{eqnarray}
\label{current2} J^{NC}_\mu=\left[\frac{1}{2}+
\frac{v^2}{2\Lambda^2}(-\alpha_j+\alpha^\prime_j)\right] \bar
{{\nu}_j}_{L}\gamma_\mu {{\nu}_j}_{L}+\left[-\frac{1}{2}+
\sin^2\theta_W-\frac{v^2}{2\Lambda^2}(\alpha_j+\alpha^\prime_j)\right]
\bar {{\ell}_j}_{L} \gamma_\mu {\ell_j}_{L}
\end{eqnarray}
In this effective current subscript "L" represents the left-handed leptons
and $v$ represents the vacuum expectation value of the scalar
field. (For definiteness, we take $v=246$ GeV and $\Lambda = 1$ TeV
in the calculations presented in this paper).

As can be seen from the current in Eq. (\ref{current2}), the
operators of Eq. (\ref{eop1}) and (\ref{eop2}) modify not only the
neutrino currents, but also the left-handed charged lepton currents.
On the other hand, right-handed charged lepton currents are not
modified. $\alpha^\prime_j$ couplings contribute both to the charged and
neutral currents but $\alpha_j$ contribute only to the neutral current.
Therefore studying charged current processes one can isolate the
couplings $\alpha^\prime_j$.
The parameters $\Delta_j$ and $\Delta_j^\prime$ introduced in the
introduction section are then expressed as follows
\begin{eqnarray}
\Delta_j=\frac{v^2}{\Lambda^2}(-\alpha_j+\alpha^\prime_j),
\,\,\,\,\,\,\,\,\,\,\,\,
 \Delta_j^\prime=2\alpha^\prime_j\frac{v^2}{\Lambda^2} .
\end{eqnarray}

\section{Neutrino fluxes and cross sections}

Accelerating $\beta$-unstable heavy ions to a given $\gamma$ factor
and allowing them to decay in the straight section of a storage
ring, very intense neutrino or anti-neutrino beams can be produced.
In the ion rest frame the neutrino spectrum is given by
\begin{eqnarray}
\frac{dN}{d\cos\theta dE_\nu}\sim
E_\nu^2(E_0-E_\nu)\sqrt{(E_\nu-E_0)^2-m_e^2}
\end{eqnarray}
where $E_0$ is the electron end-point energy, $m_e$ is the electron
mass. $E_\nu$ and $\theta$ are the energy and polar angle of the
neutrino.  The neutrino flux from accelerated ions can be obtained
by performing a boost. The neutrino flux per solid angle in a
detector located at a distance $L$ is then
\cite{BurguetCastell:2003vv}
\begin{eqnarray}
\left(\frac{d\phi^{Lab}}{dSdy}\right)_{\theta \simeq0}\simeq
\frac{N_\beta}{\pi
L^2}\frac{\gamma^2}{g(y_e)}y^2(1-y)\sqrt{(1-y)^2-y_e^2} ,
\end{eqnarray}
where $0\leq y\leq1-y_e$, $y=\frac{E_\nu}{2\gamma E_0}$,
$y_e=\frac{m_e}{E_0}$  and
\begin{eqnarray}
g(y_e)=\frac{1}{60}\left(\sqrt{1-y_e^2}(2-9y_e^2-8y_e^4)+15y_e^4Log\left[\frac{y_e}{1-\sqrt{1-y_e^2}}\right]
\right) .
\end{eqnarray}

$^{18}Ne$ and $^6He$ have been proposed as ideal candidates
for a neutrino and an anti-neutrino source, respectively
\cite{Zucchelli:2002sa,BurguetCastell:2003vv}. They produce pure
(anti-)neutrino beams via the reactions
$^{18}_{10}Ne\to^{18}_9Fe^+\nu_e$ and $^6_2He^{++}\to^6_3Li^{+++}e^-
\bar{\nu}_e$. We assume that total number of ion decays per year is
$N_\beta=1.1\times10^{18}$ for $^{18}Ne$ and
$N_\beta=2.9\times10^{18}$ for $^6He$.

\begin{figure}
\includegraphics{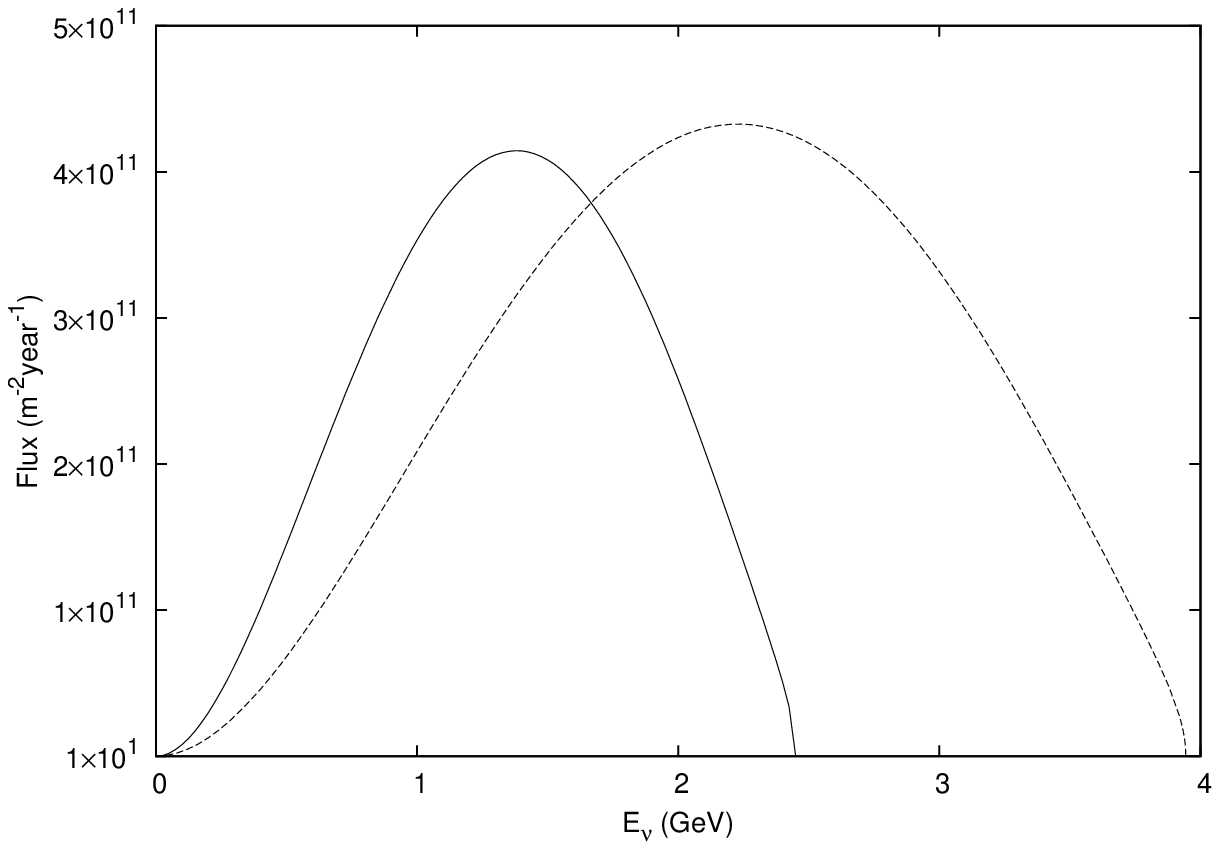}
\caption{Beta-beam fluxes as a function of neutrino energy for $\bar{\nu}_e$
(solid line) and $\nu_e$ (dotted line). $\gamma$ parameter is taken
to be 350 for $\bar{\nu}_e$ and 580 for $\nu_e$.\label{fig1}}
\end{figure}

In Fig. \ref{fig1} we plot neutrino and anti-neutrino fluxes as a
function of (anti-)neutrino energy at a detector of L = 732 km
distance. $\gamma$ parameters for ions are taken to be $\gamma=350$
for $^6He$ and $\gamma=580$ for $^{18}Ne$. The foregoing detector
distance and $\gamma$ values have been proposed in
Ref. \cite{BurguetCastell:2003vv} as a medium energy setup. In
Ref. \cite{BurguetCastell:2003vv} authors have considered a
Megaton-class water Cerenkov detector with a fiducial mass of 400
kiloton. They show that a cut demanding the reconstructed
energy to be larger than 500 MeV suppresses most of the residual
backgrounds. We assumed a water Cerenkov detector with the same mass
and a cut of 500 MeV for the calculations presented here.

We see from Fig. \ref{fig1} that neutrino spectra extend up to 4 GeV
and anti-neutrino spectra extend up to 2.5 GeV. Between 0.5 - 1.5
GeV quasi elastic nucleon scattering dominates the cross section. In
this energy range, protons scattered via inverse $\beta$-decay are
generally below Cerenkov threshold and thus it is very difficult to
discern quasi elastic scattering from neutrino-electron scattering.
Therefore we will add number of events provided by these reactions
during statistical analysis. As the energy increases, deep inelastic
scattering starts dominating the cross section. The turn-over region
is about 1.5 GeV.

\subsection{Neutrino electron scattering and neutrino nucleon quasi elastic scattering}

Electron-neutrino electron scattering in SM is described by two tree-level
diagrams containing W and Z exchange. As we have discussed in the
previous section, not only the $\nu_e\nu_e Z$ and $\nu_e eW$
vertices but also the $e^-e^-Z$ vertex is modified by the effective
Lagrangian. The total cross section is given by
\begin{eqnarray}
\label{electronscattering}
 \sigma(\nu_e e^- \to \nu_e
 e^-)=\frac{G_F^2E_\nu^2m_e}{\pi(2E_\nu+m_e)^3}\left[
 \frac{16}{3}({g_A''}^2+{g_V''}^2+g_A''g_V'')E_\nu^2
 +4m_e(2{g_A''}^2+{g_V''}^2 \right. \nonumber \\ \left.
 +g_A''g_V'')E_\nu+m_e^2({3g_A''}^2+{g_V''}^2)\right]
\end{eqnarray}
where $E_\nu$ is the initial neutrino energy, $m_e$ is the mass of
the electron and $G_F$ is the Fermi constant. The couplings $g_A''$
and $g_V''$ are defined as follows
\begin{eqnarray}
\label{couplel}
g_{A(V)}''&&=\left(1+\frac{v^2}{\Lambda^2}(-\alpha_e+\alpha^\prime_e)\right)g_{A(V)}'
+\left(1+\frac{2v^2}{\Lambda^2}\alpha^\prime_e\right)^2, \nonumber\\
g_{A(V)}'&&=g_{A(V)}-\frac{v^2}{2\Lambda^2}(\alpha_e+\alpha^\prime_e),\nonumber\\
g_{A}&&=-\frac{1}{2},\,\,\,\,\,\, g_{V}=-\frac{1}{2}+2\sin^2\theta_W ,
\end{eqnarray}
where $\Lambda$ is the energy scale of new physics and $v$ is the
vacuum expectation value of the scalar field. Anti-neutrino cross
section can be obtained from (\ref{electronscattering}) by making
the substitution $g_A'' \rightarrow -g_A''$.

As we have discussed it is very difficult to discern neutrino
electron scattering from quasi elastic scattering with a Cerenkov
detector. The differential cross section for $\nu_e n\to p \,e^-$ is
given by
\begin{eqnarray}
\label{quasielastic}
\frac{d\sigma}{d|q^2|}=&&\frac{G_F^2\cos^2\theta_C}{4\pi}
\left(1+\frac{2v^2}{\Lambda^2}\alpha^\prime_e\right)^2
\left\{(F_V+F_W+F_A)^2+(F_V+F_W-F_A)^2\left(1+\frac{q^2}{2E_\nu
m_N}\right)^2 \nonumber \right. \\ &&\left.
+\left[F_A^2-(F_V+F_W)^2\right]\frac{(-q^2)}{2E_\nu^2}+\left[F_W^2\frac{(-q^2+4m_N^2)}{4m_N^2}
-2(F_V+F_W)F_W\right] \nonumber \right. \\ &&\left.
\times\left[2+\frac{q^2(m_N+2E_\nu)}{2E_\nu^2m_N}\right]\right\}
\end{eqnarray}
where $\cos \theta_C=0.974$ is the Cabibbo angle, and $F$'s are
invariant form factors that depend on the transferred momentum
$q^2\equiv(p_p-p_n)^2$. In (\ref{quasielastic}) we ignore the terms
proportional to electron mass squared which give only a minor
contribution. The $F$'s are known as vector $F_V$, axial-vector
$F_A$ and tensor $F_W$ (or weak magnetism) form factors. They are
all G-parity invariant. We adopt the same parameterization of the
momentum dependence as in Ref. \cite{Balantekin:2006ga}:
\begin{eqnarray}
F_V(q^2)=\left(1-\frac{q^2}{(0.84\,GeV)^2}\right)^{-2} \nonumber\\
F_W(q^2)=\left(\frac{\mu_p-\mu_n}{2m_N}\right)F_V(q^2)\\
F_A(q^2)=1.262\left(1-\frac{q^2}{(1.032\,GeV)^2}\right)^{-2}\nonumber
\end{eqnarray}
Here $\mu_p-\mu_n=3.706$ is the difference in the anomalous magnetic
moments of the nucleons. We see from (\ref{quasielastic}) that quasi
elastic scattering isolates the coupling $\alpha^\prime_e$ and new
physics contribution can be factorized in the cross section.
Differential cross section for reaction ${\bar \nu}_e p\to n \,e^+$
can be obtained from (\ref{quasielastic}) by making the substitution
$F_A \rightarrow -F_A$.

We studied $95\%$ C.L. bounds using two-parameter $\chi^{2}$
analysis with and without a systematic error. The $\chi^{2}$
function is given by,
\begin{eqnarray}
\chi^{2}=\left(\frac{N_{SM}-N_{AN}}{N_{SM} \,\,
\delta_{exp}}\right)^{2}
\end{eqnarray}
where $N_{SM}$ is the number of events expected in the SM and
$N_{AN}$ is the number of events containing new physics effects. The
experimental error is
$\delta_{exp}=\sqrt{\delta_{stat}^2+\delta_{syst}^2}$ where
$\delta_{stat}$ and $\delta_{syst}$ are the statistical and
systematic errors, respectively.

In the quasi elastic scattering the main source of uncertainties
comes from the $q^2$-dependence of the form factors. The slope of
the electromagnetic form factors at $q^2=0$ is conventionally
expressed in terms of a nucleon radius. The uncertainty for these
radii is calculated to be $1\%$ \cite{Mergell:1995bf}. From this
uncertainty we have calculated the uncertainties in the number of
events. Uncertainties in the number of events for neutrino-nucleon
and antineutrino-nucleon quasi elastic scatterings are $1.1\%$ and
$0.25\%$ respectively.

In Fig. \ref{fig2} we plot $95\%$ C.L. bounds on the
$\alpha_e - \alpha^\prime_e$ parameter space
for $\nu_e$  and $\bar {\nu}_e$
scatterings. Number of events has been obtained by integrating
cross section over the (anti-)neutrino energy spectrum and
multiplying by the appropriate factor that accounts for the number of
corresponding particles (electrons, protons or neutrons) in a 400
kiloton fiducial mass of the detector. Integration ranges are $0.5 -
1.5$ GeV for quasi elastic scattering  and $0.5 - 4 (2.5)$ GeV for
$\nu_e$ $(\bar \nu_e)$ electron scattering. Number of events
provided by (anti-)neutrino electron and (anti-)neutrino nucleon
quasi elastic scatterings have been combined. We see from
Fig. \ref{fig2} that although the cross sections for $\bar {\nu}_e$
scatterings are smaller than $\nu_e$ scatterings, limits on
$\alpha_e - \alpha^\prime_e$ are almost the same. This is reasonable
since the $\bar {\nu}_e$ flux peaks at  about 1.4 GeV and it is
larger than $\nu_e$ flux everywhere in the interval $0.5 - 1.5$ GeV
(Fig. \ref{fig1}).

\subsection{Neutral- and charged-current deep inelastic scatterings}

When neutrino energy exceeds 1.5 GeV, deep inelastic scattering starts
to dominate the cross section. Since neutrino spectra extend up to 4
GeV and the deep inelastic cross sections for $\nu_e$ scattering at
this energy range are large, medium energy setup $\beta$-beam
experiment will provide high statistics deep inelastic scattering
from the nuclei. On the other hand, $\bar {\nu}_e$ deep inelastic
cross sections are smaller than the $\nu_e$ cross sections. Moreover
$\bar {\nu}_e$ spectra extend only up to 2.5 GeV and it decreases
rapidly after 1.5 GeV (Fig. \ref{fig1}). Therefore number of deep
inelastic events for anti-neutrinos is low and its statistics is
poor. So we do not perform a statistical analysis for
anti-neutrinos.

Neutral- and charged-current deep inelastic scatterings of
electron-neutrinos from the nuclei are described by t-channel Z and
W exchange diagrams respectively.
 Since quark couplings to W and Z boson are not
modified by operators (\ref{eop1},\ref{eop2}) hadron tensor does not
receive any contribution. It is defined in the standard form
\cite{Blumlein:1996vs,Forte:2001ph}
\begin{eqnarray}
W_{\mu\nu}=\left(-g_{\mu\nu}+\frac{q_\mu
q_\nu}{q^2}\right)F_1(x,Q^2)+\frac{\hat{p}_\mu \hat{p}_\nu}{p\cdot
q}F_2(x,Q^2)-i\epsilon_{\mu\nu\alpha\beta} \frac{q^\alpha
p^\beta}{2p\cdot q} F_3(x,Q^2)
\end{eqnarray}
where $p_\mu$ is the nucleon momentum, $q_\mu$ is the momentum of
the gauge boson propagator, $Q^2=-q^2$, $x=\frac{Q^2}{2p\cdot q}$
and
\begin{eqnarray}
\hat{p}_\mu\equiv p_\mu-\frac{p\cdot q}{q^2}q_\mu . \nonumber
\end{eqnarray}
The structure functions for an isoscalar target are defined as
follows \cite{CHARM1988}
\begin{eqnarray}
\label{NC}
F_2^{NC}=&&x\left[(u_L^2+u_R^2+d_L^2+d_R^2)(q_{val}+2\bar
{q})-2(u_L^2+u_R^2-d_L^2-d_R^2)(s-c)\right]\nonumber\\
F_3^{NC}=&&(u_L^2-u_R^2+d_L^2-d_R^2)q_{val}
\end{eqnarray}
\begin{eqnarray}
\label{CC}
F_2^{CC}=&&x(q_{val}+2\bar {q})+x(s-c)\nonumber\\
F_3^{CC}=&&q_{val}
\end{eqnarray}
where superscripts "NC" and "CC" represents neutral current and
charged current form factors, $q_{val}$'s are valence quark and
$q$'s are sea quark distributions. We assumed that sea quark and antiquark
distributions are the same, i.e. $q = \bar{q}$. u's and d's are defined by
\begin{eqnarray}
u_L=\frac{1}{2}-\frac{2}{3}\sin^2\theta_W,\,\,\,\,\,\,\,\,\,\,u_R=-\frac{2}{3}\sin^2\theta_W
\nonumber\\
d_L=-\frac{1}{2}+\frac{1}{3}\sin^2\theta_W,\,\,\,\,\,\,\,\,\,\,d_R=\frac{1}{3}\sin^2\theta_W\nonumber
\end{eqnarray}

The form factors $F_1$'s can be obtained from (\ref{NC}) and
(\ref{CC}) by using Callan-Gross relation $2xF_1=F_2$
\cite{Callan:1969uq}. In our calculations parton distribution
functions of Martin, Roberts, Stirling and Thorne (MRST2004)
\cite{Martin:2004dh} have been used. In our calculations we
assumed an isoscalar oxygen nucleus $N=(p+n)/2$ and two free protons
for each $H_2O$ molecule. Naturally occurring oxygen is 99.8\%
$^{16}$O which is isoscalar \cite{ti}. Hence the error incurred by
assuming an isoscalar oxygen target
would be not more than a fraction of one percent.

Possible new physics contributions coming from the operators in
(\ref{eop1}) and (\ref{eop2}) only modify the lepton tensors:
\begin{eqnarray}
\label{LeptonTensor1}
L_{\mu\nu}^{NC}=&&4\left(1+\frac{v^2}{\Lambda^2}(
-\alpha_e+\alpha_e^\prime)\right)^2 \left(k_\mu
k_\nu^\prime+k_\mu^\prime k_\nu-k\cdot k^\prime
g_{\mu\nu}+i\epsilon_{\mu\nu\alpha\beta}k^\alpha k'^\beta \right)\\
\label{LeptonTensor2}
L_{\mu\nu}^{CC}=&&8\left(1+\frac{2v^2}{\Lambda^2}\,\alpha_e^\prime\right)^2
\left(k_\mu k_\nu^\prime+k_\mu^\prime k_\nu-k\cdot k^\prime
g_{\mu\nu}+i\epsilon_{\mu\nu\alpha\beta}k^\alpha k'^\beta \right)
\end{eqnarray}
where $k_\mu$ and $k_\mu^\prime$ are the momenta of initial $\nu_e$
and final $\nu_e$ or $e^-$, respectively.

\begin{figure}
\includegraphics{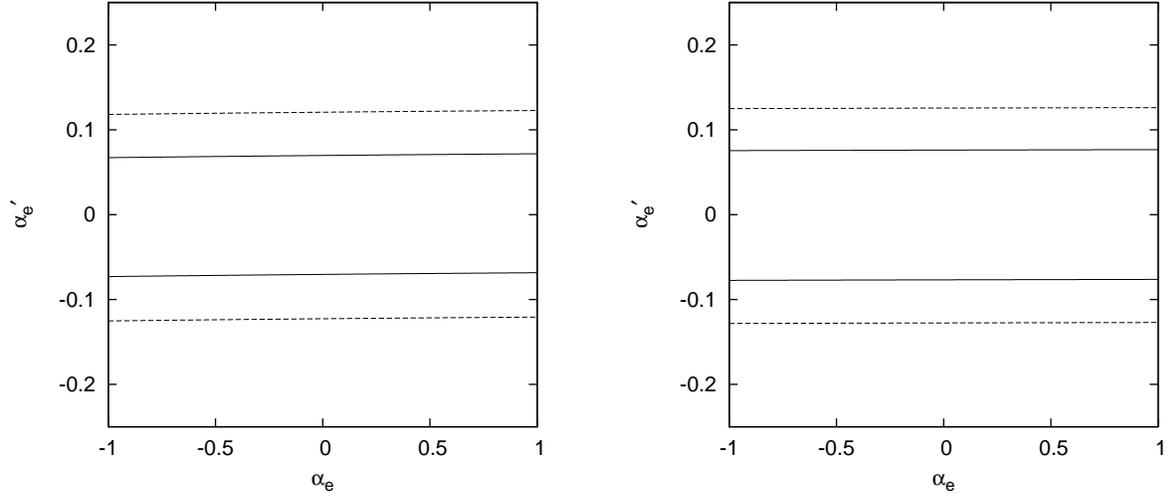}
\caption{$95\%$ C.L. sensitivity bounds on the parameter space
$\alpha_e - \alpha^\prime_e$ for $\nu_e$ (on the left) and $\bar
{\nu}_e$ (on the right) scatterings. The areas restricted by the
solid lines show the sensitivity bounds without a systematic error
and dotted lines show the sensitivity bounds with a systematic error
of 1\%. Number of events provided by (anti-)neutrino electron and
(anti-)neutrino nucleon quasi elastic scatterings have been
combined. The energy scale of new physics is taken to be $\Lambda=1$
TeV.\label{fig2}}
\end{figure}

\begin{figure}
\includegraphics{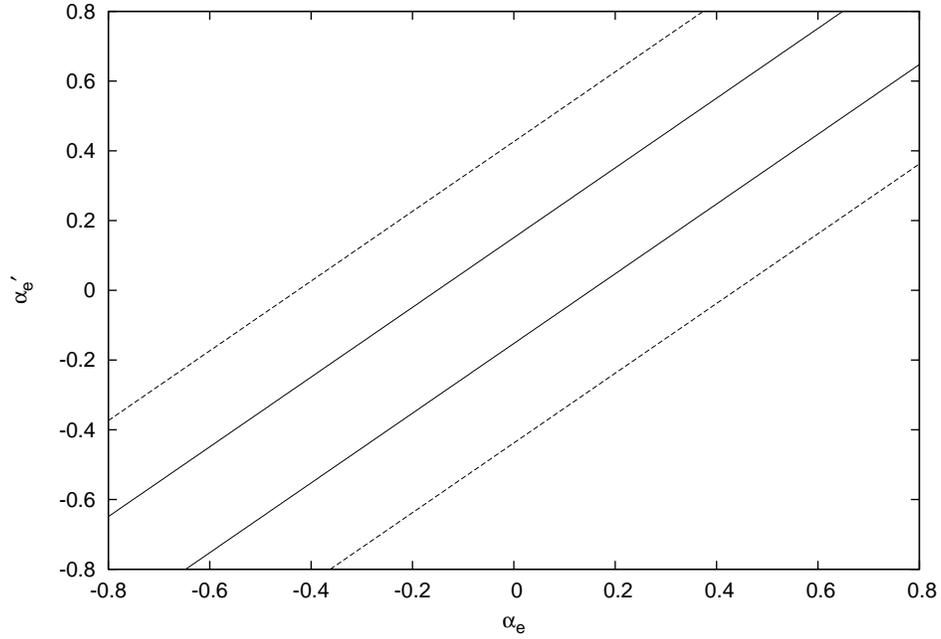}
\caption{$95\%$ C.L. sensitivity bounds on the parameter space
$\alpha_e - \alpha^\prime_e$ for neutral-current deep inelastic
scattering of $\nu_e$. The area restricted by the solid lines shows
the sensitivity bound without a systematic error and dotted lines
shows the sensitivity bound with a systematic error of 2\%. The
energy scale of new physics is taken to be $\Lambda=1$
TeV.\label{fig3}}
\end{figure}

In Fig. \ref{fig3} we show $95\%$ C.L. sensitivity bounds on the
parameter space $\alpha_e - \alpha^\prime_e$ for neutral current
deep inelastic $\nu_e$ scattering reaction. When we compare these
bounds with the bounds shown in Fig. \ref{fig2} we observe that limit
on $\alpha^\prime_e$ shown in Fig. \ref{fig3} is not as restrictive as the limit in
Fig. \ref{fig2}. For example when $\alpha_e = 0$ the limit on
$\alpha^\prime_e$ without a systematic error is
$-0.07\leq\alpha^\prime_e\leq0.07$ in Fig. \ref{fig2} (left panel).
But same limit observed from Fig. \ref{fig3} is
$-0.15\leq\alpha^\prime_e\leq0.15$. On the other hand limits on
$\alpha_e$ are very weak in Fig. \ref{fig2} as compared with
Fig. \ref{fig3}. This originates from the fact that, unlike the
neutral current deep inelastic scattering, quasi elastic scattering,
which dominates the cross section in the energy interval 0.5 - 1.5
GeV, does not contain any new physics contribution proportional to
the coupling $\alpha_e$.

\begin{figure}
\includegraphics{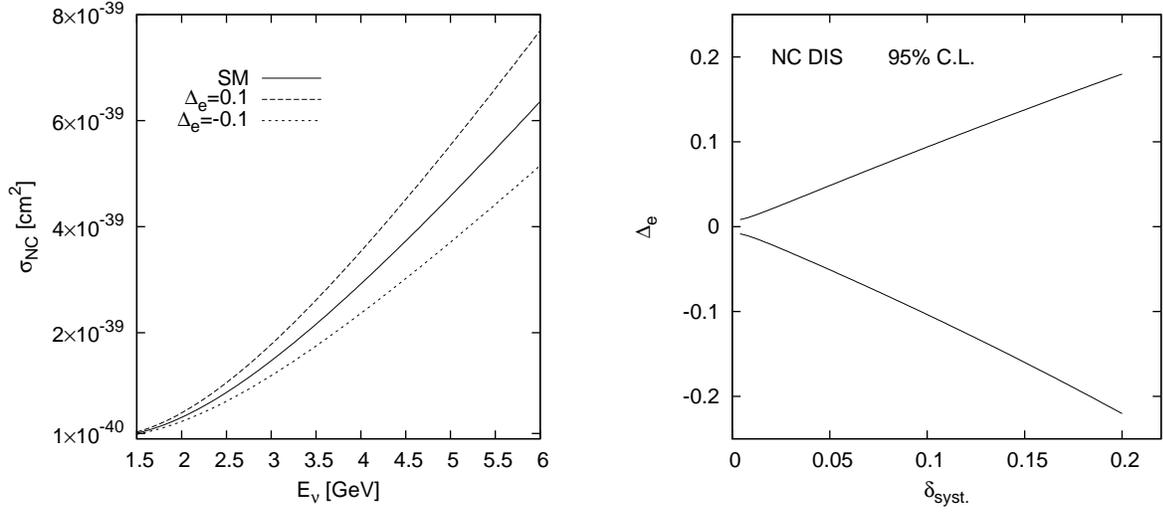}
\caption{Figure on the left shows neutral current deep inelastic
scattering cross section of $\nu_e$ from an isoscalar nucleus as a
function of neutrino energy. The legends are for standard model (SM)
and various values of the anomalous coupling
$\Delta_e=\frac{v^2}{\Lambda^2}( -\alpha_e+\alpha_e^\prime)$. Figure
on the right shows 95\% C.L. limits on $\Delta_e$ as a function of
systematic error. The energy scale of new physics is taken to be
$\Lambda=1$ TeV.\label{fig4}}
\end{figure}

\begin{figure}
\includegraphics{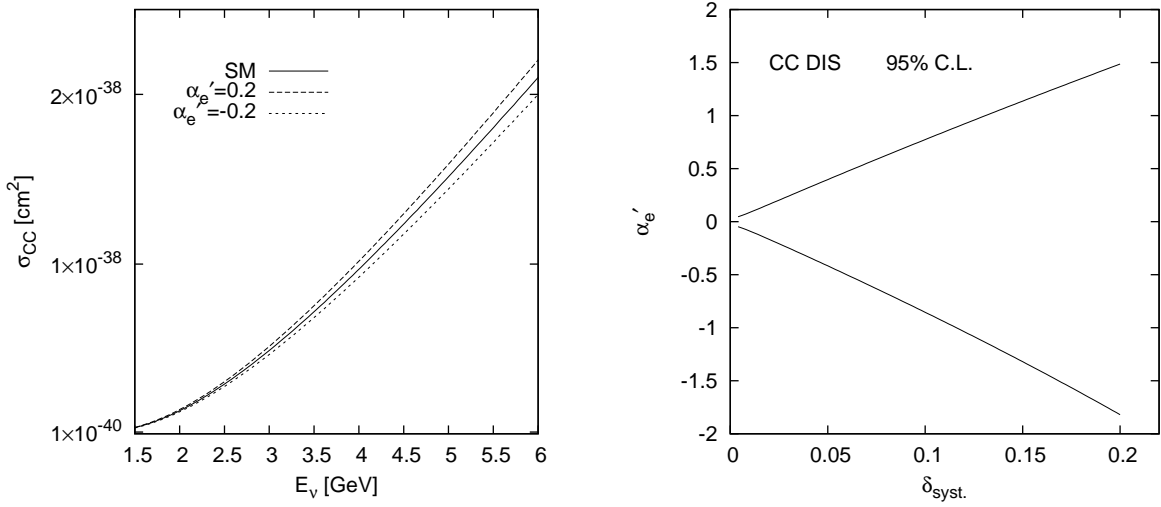}
\caption{Figure on the left shows charged current deep inelastic
scattering cross section of $\nu_e$ from an isoscalar nucleus as a
function of neutrino energy. The legends are for standard model (SM)
and various values of the anomalous coupling $\alpha^\prime_e$.
Figure on the right shows 95\% C.L. limits on $\alpha^\prime_e$ as a
function of systematic error. The energy scale of new physics is
taken to be $\Lambda=1$ TeV.\label{fig5}}
\end{figure}

The behavior of the neutral (charged) current deep inelastic
scattering cross section as a function of initial neutrino energy is
plotted for various values of the anomalous coupling $\Delta_e$
($\alpha^\prime_e$) in the left panel of Fig. \ref{fig4} (Fig.
\ref{fig5}). We see from these figures that deviation of the
anomalous cross sections from their SM values increases in magnitude
as the energy increases. On the other hand, the percentage change in the
cross section is energy independent.  This is clear from the energy
independence of new physics contributions. However the cross sections
and therefore the statistics increase with the energy. We see from
the figures that the increment in the cross sections is linear
approximately after 3.5 GeV. Therefore high energy neutrino
experiments are expected to reach a high sensitivity to probe these
anomalous couplings. 95\% C.L. limits on anomalous couplings
$\Delta_e$ and $\alpha^\prime_e$ are plotted as a function of
systematic error for neutral and charged current deep inelastic
scattering processes in the right panels of Fig. \ref{fig4} and Fig.
\ref{fig5}. $95\%$ C.L. sensitivity bounds on $\Delta_e$ and
$\alpha^\prime_e$ are $-0.02\leq\Delta_e\leq0.02$ and
$-0.167\leq\alpha^\prime_e\leq0.164$ with a systematic error of
$2\%$. These bounds can be compared with CHARM and LEP limits
(\ref{CHARMlimit}) and (\ref{LEPlimit}). We see that medium energy
setup of the $\beta$-beam experiment with 1 year of running and a
systematic error of $2\%$ provides approximately 10 times more
restricted limit for $\Delta_e$ as compared with the CHARM limit.
This limit is 4 times more restricted even systematic error is
$5\%$. On the other hand, limit on $\alpha^\prime_e$ with a
systematic error of $2\%$ is approximately 1.3 times worse than the
LEP limit.

It is important to discuss uncertainties on these couplings due to
uncertainties from structure functions and SM electroweak
parameters. During calculations we used the following values for
some SM parameters: $G_F = 1.16637(1) 10^{-5} GeV^{-2}$,
$\sin^2\theta_W=0.23122(15)$, $\sin\theta_C=0.227(1)$
\cite{Yao:2006px}. Here numbers in parentheses after the values give
1-standard-deviation uncertainties in the last digits. Uncertainty
on the limit of $\Delta_e$ in neutral current deep inelastic
scattering due to uncertainties from the above SM parameters is
order of $10^{-5}$.  Uncertainty on the limit of $\alpha^\prime_e$
is order of $10^{-6}$ in charged current deep inelastic scattering
and order of $10^{-5}$ in the combined analysis of (anti-)neutrino
electron and (anti-)neutrino nucleon quasi elastic scatterings.
Uncertainties in the structure functions may lead to a considerable
uncertainty in the cross sections. Nucleon structure functions were
precisely measured in neutrino-iron and anti-neutrino-iron
scattering reactions at the Fermilab Tevatron by the CCFR
collaboration. The systematic error of $2.1\%$ was reported in the
cross sections \cite{Seligman}. In the near future, the precision on
the structure functions is expected to increase dramatically
\cite{Martin:2002aw}. In this context beta beam facility itself can
be used to reduce uncertainties in the structure functions. Beta
beams present an ideal venue to measure neutrino cross sections.
For beta beams neutrino fluxes are precisely known and therefore uncertainties
associated with the neutrino (anti-neutrino) fluxes are negligible.
The Lorentz $\gamma$ factor of the accelerated ions can be varied.
We see from (\ref{LeptonTensor1}) and (\ref{LeptonTensor2}) that new
physics contributions are factorized in the cross sections.
Therefore, the ratio of deep inelastic cross sections measured in
two different $\gamma$ factors is independent from the new physics
contributions that we considered. Theoretical predictions can
be fitted to the measured ratio in order to eliminate uncertainties.
This procedure can also be done for the ratio of neutrino and
anti-neutrino deep inelastic cross sections. The ratio of neutrino
and anti-neutrino deep inelastic cross sections is again independent
from the new physics contributions and can be especially used to
reduce the uncertainty in the structure function $F_3$.

\subsection{Different $\gamma$ options}

\begin{figure}
\includegraphics{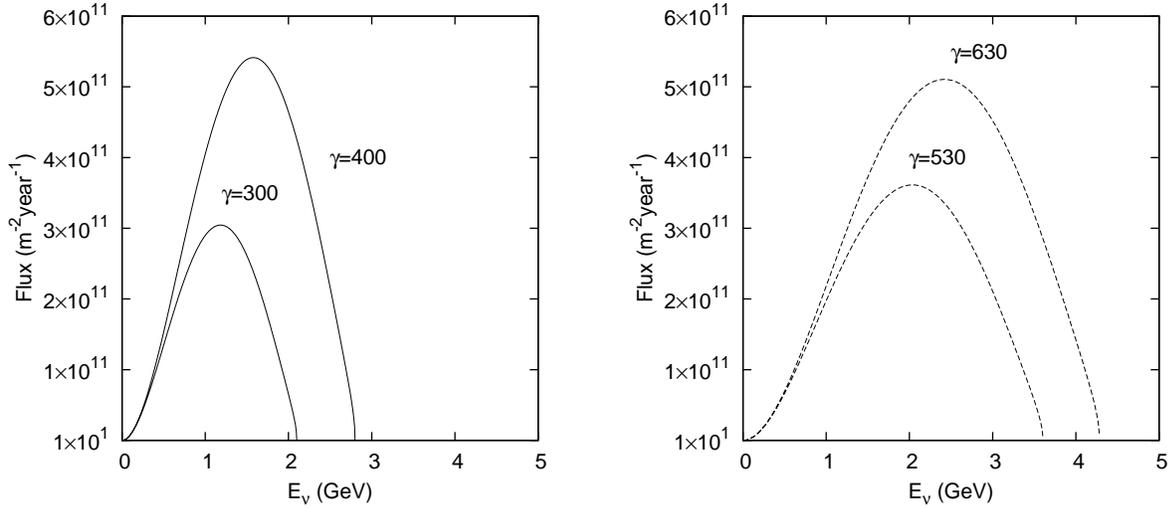}
\caption{Fluxes as a function of neutrino energy for different
values of the parameter $\gamma$ stated on the figures. Figure on
the left (right) shows fluxes for $\bar{\nu}_e$
($\nu_e$).\label{fig6}}
\end{figure}

\begin{figure}
\includegraphics{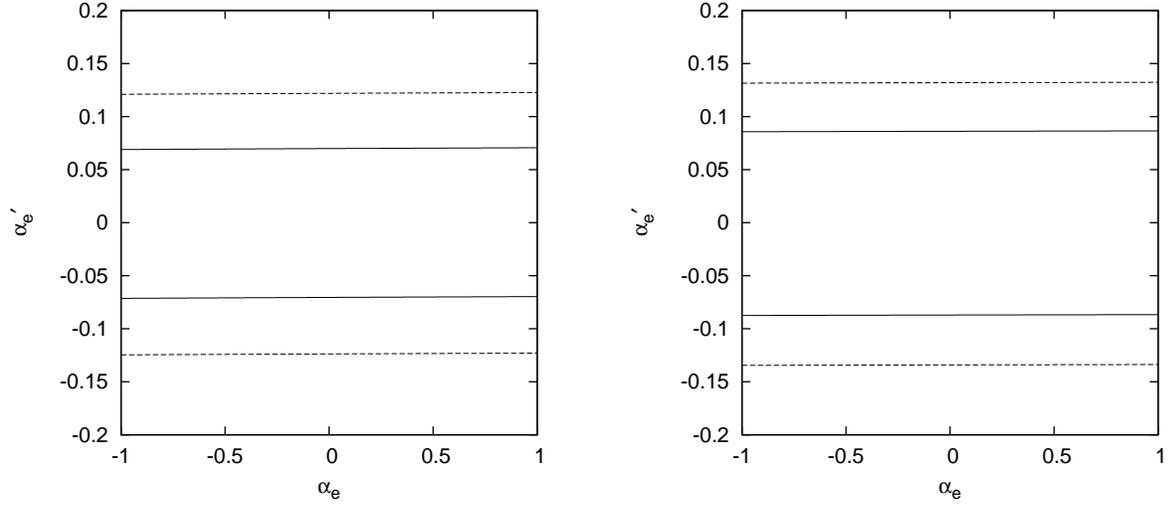}
\caption{$95\%$ C.L. sensitivity bounds on the parameter space
$\alpha_e - \alpha^\prime_e$ for $\bar \nu_e$ scattering. Left panel
is for $\gamma$=400 and right panel is for $\gamma$=300.  The areas
restricted by the solid lines show the sensitivity bounds without a
systematic error and dotted lines show the sensitivity bounds with a
systematic error of 1\%. Number of events provided by anti-neutrino
electron and anti-neutrino nucleon quasi elastic scatterings have
been combined. The energy scale of new physics is taken to be
$\Lambda=1$ TeV.\label{fig7}}
\end{figure}

It is important to investigate the variation of the sensitivity
limits when the $\gamma$ parameter of the ion beams are changed.
Different from the proposed $\gamma$ values for a medium energy
setup in Ref.\cite{BurguetCastell:2003vv} we consider $\gamma = 300$
and 400 for $^6He$ and $\gamma$ = 530 and 630 for $^{18}Ne$. The
fluxes for these $\gamma$ values at a detector of 732 km distance
are plotted in Fig. \ref{fig6}. We see from this figure that $\nu_e$
fluxes in the energy interval 0 - 1.5 GeV change very slightly with
$\gamma$. Therefore the combined statistics of neutrino quasi
elastic  and neutrino electron scatterings do not change
significantly.  On the other hand, $\bar {\nu}_e$ fluxes rapidly
change after 1 GeV with $\gamma$. Combined limits of anti-neutrino
electron and anti-neutrino nucleon quasi elastic scatterings for
$\gamma$=400 and $\gamma$=300 are given in Fig. \ref{fig7}.

\begin{figure}
\includegraphics{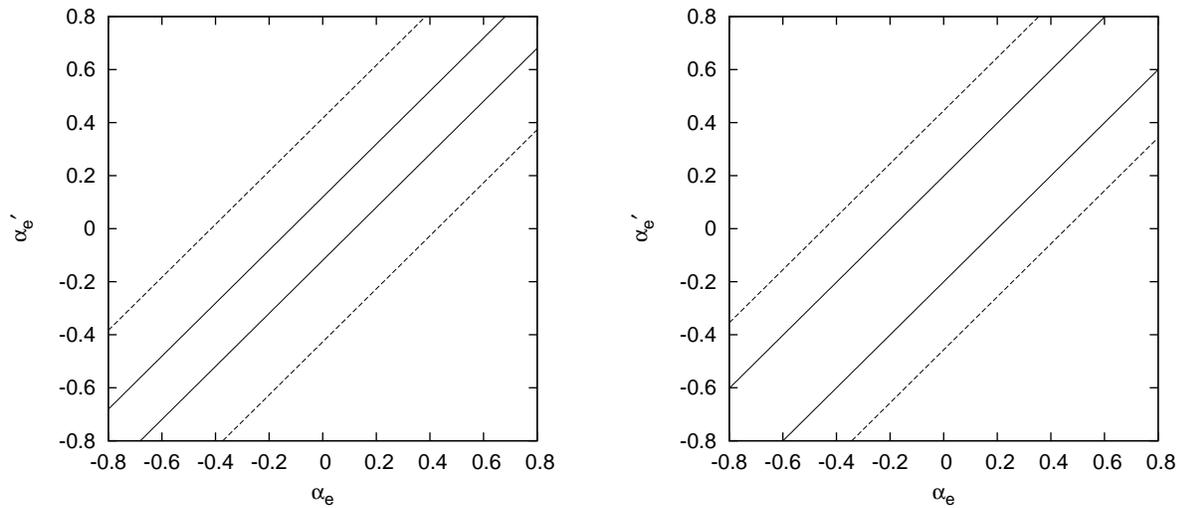}
\caption{$95\%$ C.L. sensitivity bounds on the parameter space
$\alpha_e - \alpha^\prime_e$ for neutral current deep inelastic
scattering of $\nu_e$. Left panel is for $\gamma$=630 and right
panel is for $\gamma$=530.  The area restricted by the solid lines
shows the sensitivity bound without a systematic error and dotted
lines shows the sensitivity bound with a systematic error of 2\%.
The energy scale of new physics is taken to be $\Lambda=1$
TeV.\label{fig8}}
\end{figure}

\begin{figure}
\includegraphics{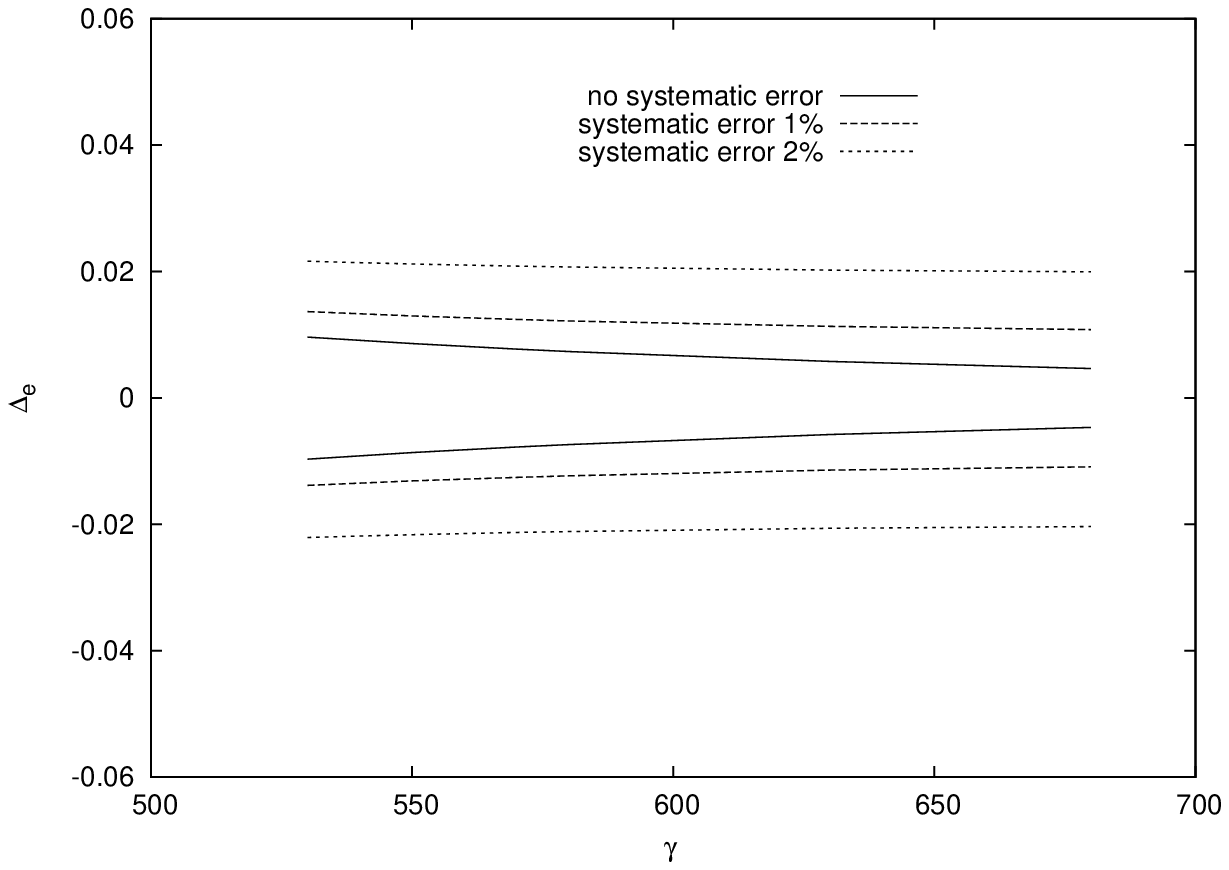}
\caption{95$\%$ C.L. bounds on $\Delta_e$ as a function of Lorentz
$\gamma$ factor with various systematic errors stated on the figure.
Bounds obtained from neutral current deep inelastic $\nu_e$
scattering. The energy scale of new physics is taken to be
$\Lambda=1$ TeV.\label{fig9}}
\end{figure}

\begin{figure}
\includegraphics{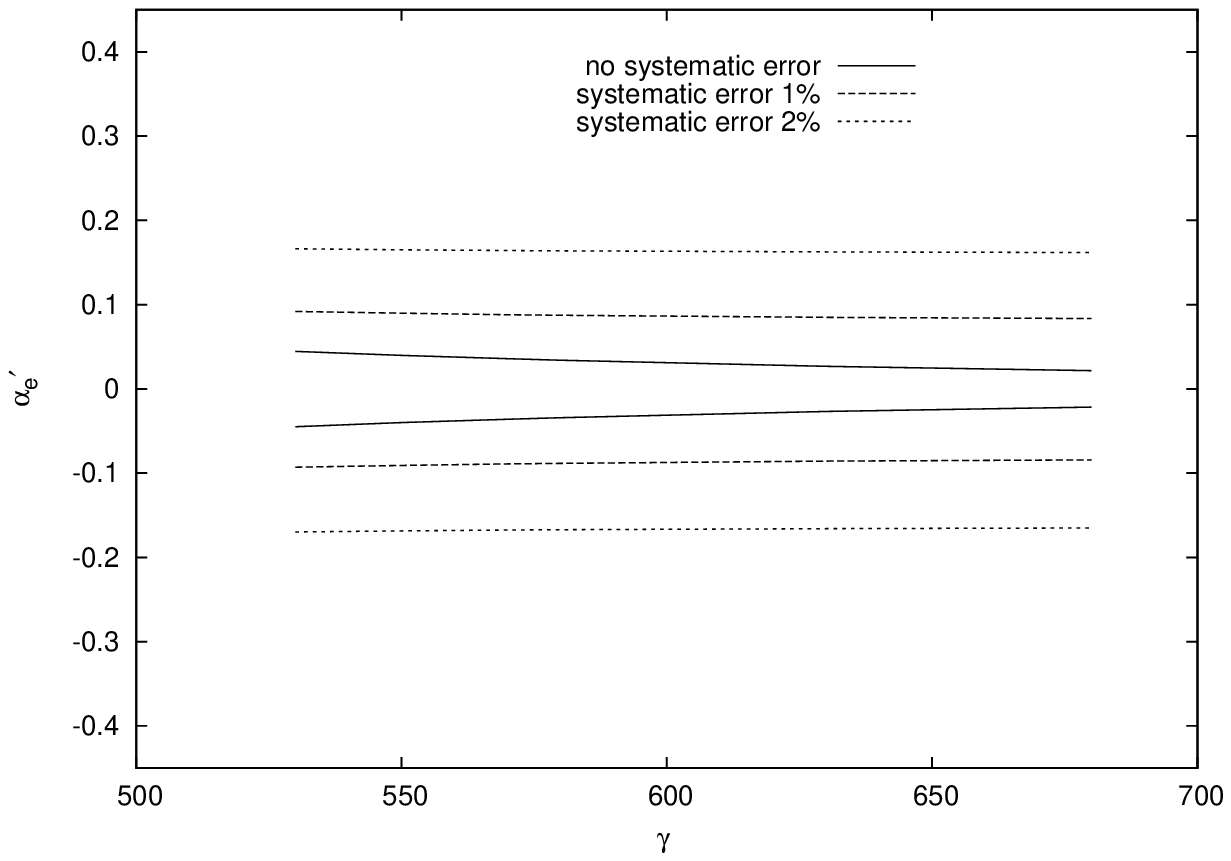}
\caption{95$\%$ C.L. bounds on $\alpha^\prime_e$ as a function of
Lorentz $\gamma$ factor with various systematic errors stated on the
figure. Bounds obtained from charged current deep inelastic $\nu_e$
scattering. The energy scale of new physics is taken to be
$\Lambda=1$ TeV.\label{fig10}}
\end{figure}

The number of deep inelastic events increase with $\gamma$ due to
two reasons:  First, energy spectra of the neutrinos extend to
higher energy values. Second, average fluxes grow with $\gamma$.
Therefore one can expect a sizable improvement in the limits as the
$\gamma$ increases. In order to compare limits for different
$\gamma$ options we present Figs. \ref{fig8}-\ref{fig10}. We see
from Fig.\ref{fig8} that limits on $\alpha_e - \alpha^\prime_e$
without a systematic error improves by more than a factor of 1.5 as
the $\gamma$ increases from 530 to 630. In Fig.\ref{fig9} and
Fig.\ref{fig10} we show the behavior of $95\%$ C.L. sensitivity
bounds as a function of Lorentz $\gamma$ factor. We see from
Fig.\ref{fig9} that $95\%$ C.L. sensitivity bounds on $\Delta_e$
with a systematic error of $2\%$ are $-0.021\leq \Delta_e \leq0.020$
for $\gamma$=630 and $-0.022\leq \Delta_e \leq0.022$ for
$\gamma$=530. The influence of $\gamma$ on the limits of the
coupling $\alpha^\prime_e$ obtained from charged current deep
inelastic scattering can be observed from Fig.\ref{fig10}. From
Fig.\ref{fig10} we have $95\%$ C.L. limits of $-0.166\leq
\alpha^\prime_e \leq0.163$ for $\gamma$=630 and $-0.170\leq
\alpha^\prime_e \leq0.166$ for $\gamma$=530 with a systematic error
of $2\%$.

\section{Conclusions}

Experiments that isolate only a single neutrino flavor such as
$\beta$-beam proposals or NuSOnG \cite{Adams:2008cm} proposal do not
require neutrino flavor universality assumption and therefore
provide more information about new physics probes on neutrino-gauge
boson couplings. In this paper, we explored signatures for deviation
from the SM predictions in neutrino-Z boson and neutrino-W boson
couplings. We do not {\it a priori} assume universality of the
couplings of neutrinos to these gauge bosons. We deduce that medium
energy setup of the $\beta$-beam experiment has a great potential to
probe possible new physics contributions to $Z\nu_e\nu_e$ coupling.
Beta beam experiment with a systematic error of $2\%$ improves the
limit on $Z\nu_e\nu_e$ approximately a factor of 10 compared with
CHARM limit. It also probes $We\nu_e$ coupling with a good
sensitivity. The limit obtained for the coupling $We\nu_e$ is in the
same order of the LEP limit. Coupled with possible complementary
measurements of muon-neutrino or/and tau-neutrino scattering cross
sections for example at NuSOnG experiment
\cite{Balantekin:2008rc,Balantekin:2008ib}, beta beam experiment can
be a powerful probe of new neutrino physics.

\begin{acknowledgments}

This work was supported in part
by the U.S. National Science Foundation Grant No.\ PHY-0555231 and
in part by the University of Wisconsin Research Committee with funds
granted by the Wisconsin Alumni Research Foundation. \.{I}.
\c{S}ahin and B. \c{S}ahin acknowledge support through the
Scientific and Technical Research Council (TUBITAK) BIDEB-2219
grant.
\end{acknowledgments}


\pagebreak

\end{document}